\documentclass[twocolumn,superscriptaddress,prl]{revtex4-1}
\usepackage[T1]{fontenc}
\usepackage[utf8]{inputenc}
\usepackage{etoolbox}
\usepackage{mathrsfs}
\usepackage{amssymb, amsbsy, amsmath, latexsym, dsfont, array, layout, graphicx,mathrsfs,color}
\usepackage{soul}
\usepackage{hyperref} \hypersetup{colorlinks=true, citecolor=blue, urlcolor=blue, linkcolor=blue}
\usepackage{mathtools}
\usepackage{ulem}

\usepackage{placeins}

\makeatletter
\newenvironment{norulewidetext}{%
  \par\ignorespaces
  \onecolumngrid
  \vskip10\p@
  \prep@math@patch
}{%
  \par
  \vskip8.5\p@
  \twocolumngrid\global\@ignoretrue
  \@endpetrue
}
\makeatother

\newcommand{\EndMatterTitle}{%
  \par\addvspace{1.2\baselineskip}%
  \begin{norulewidetext}
    \centering{\bfseries\large End Matter\par}
  \end{norulewidetext}
  \addvspace{0.8\baselineskip}%
}

\newcommand{\one}{\mathds{1}}

\DeclarePairedDelimiter{\ket}{\lvert}{\rangle}
\DeclarePairedDelimiter{\bra}{\langle}{\rvert}
\DeclarePairedDelimiterX{\braket}[2]{\langle}{\rangle}{#1 \delimsize\vert #2}

\begin{document}

\title{Parent Hamiltonian and intrinsic phase transition in non-Hermitian photonic systems}
\author{Yuntao Xiao}
\thanks{These authors contributed equally}
\affiliation{Beijing Computational Science Research Center, Beijing 100084, China}

\author{Yuchen Guo}
\thanks{These authors contributed equally}
\affiliation{State Key Laboratory of Low Dimensional Quantum Physics and Department of Physics, Tsinghua University, Beijing 100084, China}

\author{Xiaojian Huang}
\affiliation{Beijing Computational Science Research Center, Beijing 100084, China}

\author{Huixia Gao}
\affiliation{School of Physics, Southeast University, Nanjing 211189, China}

\author{Dengke Qu}
\affiliation{School of Physics, Southeast University, Nanjing 211189, China}

\author{Lei Xiao}
\affiliation{School of Physics, Southeast University, Nanjing 211189, China}

\author{Kunkun Wang}
\email{kunkunwang@126.com}
\affiliation{School of Physics and Optoelectronic Engineering, Anhui University, Hefei 230601, China}

\author{Shuo Yang}
\email{shuoyang@tsinghua.edu.cn}
\affiliation{State Key Laboratory of Low Dimensional Quantum Physics and Department of Physics, Tsinghua University, Beijing 100084, China}
\affiliation{Frontier Science Center for Quantum Information, Beijing 100084, China}
\affiliation{Hefei National Laboratory, Hefei 230088, China}

\author{Peng Xue}
\email{gnep.eux@gmail.com}
\affiliation{School of Physics, Southeast University, Nanjing 211189, China}

\begin{abstract}
Non-Hermitian systems host phenomena absent in Hermitian physics, but realizing Hamiltonians with intrinsic non-Hermitian properties remains challenging.
The theoretical method of non-Hermitian parent Hamiltonian (NH-PH) enables the construction of a non-Hermitian system from a pair of matrix product states (MPSs) with tailored properties.
Here, we report the first experimental generation of NH-PHs. This generation starts from MPSs that represent asymmetric Affleck--Kennedy--Lieb--Tasaki (AKLT) states.
The construction is validated with single photons via imaginary-time evolution of the generated NH-PH to obtain its left and right ground states.
We then characterize the properties of the system by measuring four different order parameters that probe non-reciprocal correlations, chiral imbalance, and conventional antiferromagnetic correlations.
Furthermore, extending the framework to a larger system with a different model, we observe an intrinsic non-Hermitian phase transition, manifested by abrupt jumps of an order parameter when the designated zero-energy modes cease to be the globally lowest-energy states.
Our work provides the first experimental realization and characterization of non-Hermitian Hamiltonians with controllable and customizable properties, opening new avenues for exploring intrinsic non-Hermitian phenomena across diverse physical platforms.

\end{abstract}

\maketitle

\textit{Introduction.}---
Non-Hermitian systems have attracted significant attention in recent years due to the emergence of novel phenomena absent in Hermitian systems~\cite{PhysRevLett.77.570,Bender_2007,PhysRevB.97.045106, Ashida02072020, Pan2020},
such as exceptional points~\cite{Muller_2008,Mohammad2019,Ananya2020,Ding2022,MaoWenbo2024,Xue2026} and the non-Hermitian skin effect~\cite{PhysRevLett.116.133903,PhysRevLett.121.136802,PhysRevLett.123.170401,Li2020,Zhang2021}. These phenomena have enabled a wide range of applications, including enhanced sensing~\cite{Xiao2024}, non-Hermitian engineered single-mode lasing~\cite{Li2023Laser}, robust transport~\cite{delPino2022}, chiral state transfer~\cite{Khandelwal2024,Gao2025}, and accelerated entanglement generation~\cite{Li2023Ent},
and have been experimentally observed on various platforms such as optical~\cite{Xiao2017,Chen2017,Reisenbauer2024,Dai2024,Wang2023SciAdv,Xiao2025}, acoustic~\cite{Gao2021,Zhang2021Twisted,PhysRevLett.133.126601}, superconducting circuits~\cite{PhysRevLett.127.140504}, among others~\cite{Helbig2020, LiuShuo2021,Gao2015}. However, most of the research so far follows a common strategy, i.e., introducing non-Hermitian terms by gain/loss or other dissipative features into well-established Hermitian Hamiltonians to construct a non-Hermitian system\nolinebreak[4]\mbox{~\cite{PhysRevLett.121.086803,PhysRevX.8.031079,PhysRevLett.124.056802,PhysRevLett.121.026808}}.

This naturally raises a fundamental question: can one directly design a non-Hermitian Hamiltonian with prescribed properties, rather than perturbing an existing Hermitian model?
In Hermitian systems, this is elegantly achieved through the parent Hamiltonian method~\cite{PhysRevLett.59.799,PerezGarcia2007}, which allows the construction of a local, gapped Hamiltonian with a predefined ground state represented as matrix product state (MPS)~\cite{Verstraete01032008,SCHOLLWOCK201196,ORUS2014117,RevModPhys.93.045003}.
This method, which offers significant advantages for studying the topological phases of matter, has been widely applied in various systems~\cite{PhysRevB.89.165125,PhysRevX.5.011024}.
Recently, this concept was generalized to non-Hermitian systems~\cite{PhysRevLett.130.220401}, where a non-Hermitian parent Hamiltonian (NH-PH) can be analytically constructed from a pair of MPSs serving as left and right zero-energy modes.
By prescribing specific structures in these zero modes, one can directly construct the corresponding NH-PH with desired properties~\cite{Guo2023Composite}. This construction therefore provides a general framework for systematically exploring intrinsic non-Hermitian phenomena.

\begin{figure*}[t]
    \centering
    \includegraphics[width=0.9\textwidth]{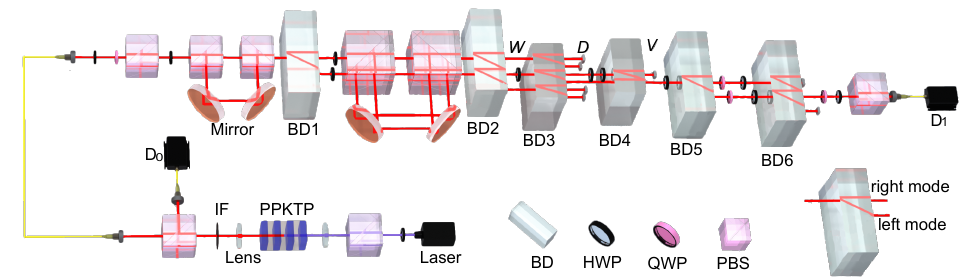}
    \caption{
    Experimental setup. A heralded single photon is created via type-II spontaneous parametric down-conversion via a periodically poled potassium titanyl phosphate crystal (PPKTP).
    In the state space spanned by the polarization modes $\ket{h},\ket{v}$ and the spatial modes $\ket{r},\ket{l}$, the signal photon is prepared in a maximally mixed state using two unbalanced interferometer setups composed of four polarizing beam splitters (PBSs), three half-wave plates (HWPs), and a beam displacer (BD1).
    We purify this mixed state into the ground states of the NH-PH via four BDs (BD2--BD5) and HWPs.
   The resulting ground states are reconstructed via quantum state tomography performed with BD6, a PBS, three quarter-wave plates (QWPs), and three HWPs.
    Finally, the photons are detected by avalanche photodiodes (APDs) and recorded by the coincidence counts of $D_0$ and $D_1$.}
     \label{fig1}
\end{figure*}

Nevertheless, how to realize and probe such NH-PH in real experiments remains an open problem, hindering its broad applications in the current form.
In this work, we report the first experimental realization and characterization of NH-PHs.
Our protocol begins by specifying desired non-Hermitian features in the biorthogonal ground states, and encoding them in a pair of predesigned MPSs that serve as the target right and left zero-energy ground states.
Using the NH-PH approach~\cite{PhysRevLett.130.220401}, we construct the Hamiltonians whose ground states are the prescribed MPS pair, thereby building in the desired properties by design.

Here, we experimentally generate and characterize NH-PHs for two models using single photons for the first time.
As a proof-of-principle validation, we first implement an $N = 2$ system based on the MPS representation of asymmetric Affleck--Kennedy--Lieb--Tasaki (AKLT) states~\cite{Maekawa2023}. By measuring four order parameters, we characterize its non-reciprocal correlations, chiral imbalance, and antiferromagnetic correlations.
To illustrate a specific application, we then extend to an $N = 3$ model that exhibits an intrinsic non-Hermitian phase transition, where the ferromagnetic order parameter shows an abrupt jump accompanied by the emergence of a negative-energy eigenmode.
Our work establishes a systematic blueprint for reverse-engineering non-Hermitian phases, and provides a versatile platform for exploring new properties and phenomena in non-Hermitian physics, opening up a new avenue toward controllable non-Hermitian quantum simulation.

\textit{Construction of the NH-PH.}---
In non-Hermitian systems,
physical observables $O$ are evaluated within a biorthogonal framework~\cite{Brody_2014}
\begin{equation}
    \langle {O} \rangle_{LR} = \frac{\langle L | {O} | R \rangle}{\langle L | R \rangle},
    \label{eq1}
\end{equation}
where \(|R\rangle\) and \(|L\rangle\) are the right and left ground states of the non-Hermitian Hamiltonian, respectively.
Thus, by predesigning \(|R\rangle\) and \(|L\rangle\) with particular properties, we can experimentally realize non-Hermitian systems with desired features.

To construct the NH-PH, we start from a pair of short-range correlated MPSs on an $N$-site chain under open boundary conditions (OBC)~\cite{SeeSM}
\begin{align}
    \ket{R(L)} &= \sum_{\tau, \nu} l_{\nu_1}
    \left( \prod_{j=1}^{N}
    [A_{j}^{R(L)}]^{\tau_j}_{\nu_j, \nu_{j+1}} \right)
    r_{\nu_{N+1}} \nonumber \\
    &\quad \times \ket{\tau_1, \tau_2, \cdots, \tau_N},
    \label{eq2}
\end{align}where $[A_{j}^{R(L)}]$ is the local tensor at the site $j$, $\tau_j$ is the physical index with dimension $d_p$, $\{\nu_j, \nu_{j+1}\}$ are virtual indices with bond dimension $d_v$, while $l$ and $r$ are boundary tensors.
Recent theoretical work~\cite{PhysRevLett.130.220401} has shown that given such a pair of MPSs, one can analytically construct an NH-PH for which $\ket{R}$ and $\ket{L}$ serve as the right and left zero-energy modes, respectively.

\begin{figure*}[t]
    \centering
    \includegraphics[width=0.9\textwidth]{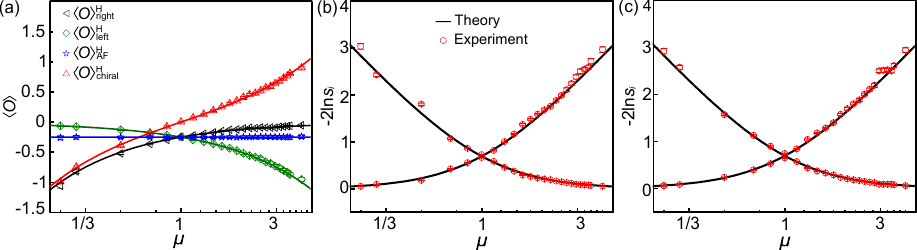}
    \caption{(a) The expectation values of order parameters $\langle O \rangle=\langle L|{O}|R\rangle / \langle L|R\rangle$ for NH-PH $H$ with $N=2$.
    Here, $\langle O^{\text{H}} \rangle$ refers to the results extracted from the reconstructed $|{\Psi^s_R}\rangle$ and $|{\Psi^s_L}\rangle$ after the imaginary-time evolution.
    The entanglement spectrum of (b) $|\Psi^s_R\rangle$ and (c) $|\Psi^s_L\rangle$ obtained from state tomography.
    Experimental data are represented by dots, and theoretical results are denoted by continuous lines.
    Error bars indicate the statistical uncertainty, which is obtained based on assuming Poissonian statistics.
    }
\label{fig2}
\end{figure*}

As a concrete illustration, we first adopt the same example as in Ref.~\cite{PhysRevLett.130.220401}, where the asymmetric AKLT state in a spin-1 qutrit system is chosen as $\ket{R}$, whose nonzero tensor elements include~\cite{Maekawa2023}
\begin{equation}
\begin{aligned}
[A_j^R]^{[0]}_{10} &= -\sqrt{\mu}, \quad
[A_j^R]^{[2]}_{01} = \sqrt{\mu}, \\
[A_j^R]^{[1]}_{00} &= \frac{1}{\sqrt{2}}, \quad
[A_j^R]^{[1]}_{11} = -\frac{\mu}{\sqrt{2}}.
\end{aligned}
\end{equation}
In the representation of valence bond solid~\cite{Wei2022, Smith2023}, the built-in chirality is encoded in the asymmetric virtual bond $|0)(1|-\mu\,|1)(0|$, characterized by the asymmetry parameter $\mu$.
We first specialize to the model with $N=2$ under the OBC of $\nu_1 = {1}$ and $\nu_{N+1} = {0}$, leading to the following tensor elements:
\begin{equation}
\begin{aligned}
[A_1^R]^{[0]}_{10} &= -\sqrt{\mu}, \quad
[A_1^R]^{[1]}_{11} = -\frac{\mu}{\sqrt{2}}, \\
[A_2^R]^{[1]}_{00} &= \frac{1}{\sqrt{2}}, \quad
[A_2^R]^{[0]}_{10} = -\sqrt{\mu}.
\end{aligned}
\end{equation}In this scenario, each local tensor only supports a dim-$2$ subspace $\{\ket{0}, \ket{1}\}$ of the original spin-$1$ degree of freedom, allowing a map to a spin-$1/2$ state.
Thus, the MPS yields the corresponding singlet state \(|R\rangle \propto |01\rangle - \mu |10\rangle\) with an asymmetric valence bond.

Meanwhile, we choose the left state as $\ket{L}=\mathcal{P}\ket{R}$ with the parity operator $\mathcal{P}$, whose tensor elements satisfy $[A_j^L]^{[\tau]} = \left([A_j^R]^{[\tau]}\right)^{\rm T}$, i.e., transpose regarding the two virtual indices.
Following the construction in Ref.~\cite{PhysRevLett.130.220401}, the resulting NH-PH for $N=2$ reads~\cite{SeeSM}
\begin{equation}
    H=
    \begin{pmatrix}
    1 & 0 & 0 & 0 \\
    0 & \frac{1}{2} & \frac{1}{2 \mu} & 0 \\
    0 & \frac{\mu}{2} & \frac{1}{2} & 0 \\
    0 & 0 & 0 & 1
    \end{pmatrix},
    \label{eq9}
\end{equation}
with eigenvalues $\{0,1,1,1\}$, where the designed $\ket{R}$ and $\ket{L}$ are the right and left ground states with zero energy, respectively.

\textit{Experimental implementation.}---
To experimentally verify that the constructed NH-PH exhibits the intended properties, we implement the imaginary-time evolution to extract both the right and left ground states of the NH-PH and reconstruct them by quantum state tomography~\cite{McArdle2019,Motta2020}.
Imaginary-time evolution exponentially suppresses higher-energy eigenstates, projecting the initial state onto the ground state for a sufficiently long evolution time.
As shown in Fig.~\ref{fig1}, the basis states of the system are encoded as
$\ket{00} = \ket{r}\ket{h},\ \ket{01} = \ket{r}\ket{v},\ \ket{10} = \ket{l}\ket{h},\ \ket{11} = \ket{l}\ket{v},$
where $\ket{r}$ and $\ket{l}$ represent the right and left spatial modes of single photons and \(\ket{h}\) and \(\ket{v}\) represent the horizontal and vertical polarizations of the photons.

Photons are first initialized in the horizontal polarization using a polarizing beam splitter (PBS). We then prepare the photons in a maximally mixed state $\rho_0=\one/4$, where $\one$ denotes the $4\times 4$ identity matrix.
This is implemented using a beam displacer (BD1) together with a set of half-wave plates (HWPs), which distribute the photons with equal probabilities among the four basis states.
The photons then pass through two unbalanced interferometers, where the path difference exceeds the coherence length, effectively destroying coherence between different basis components and yielding the maximally mixed state $\rho_0$.

To extract the right (left) ground state of the constructed NH-PH, we apply imaginary-time evolution $U_\tau=e^{-H\tau}$ ($U_\tau^\dagger=e^{-H^\dagger\tau}$) to the initial state $\rho_0$ with a fixed evolution time $\tau$.
In the eigenbasis of $H$, this evolution suppresses each eigenstate by a factor $e^{-E_j \tau}$.
Here, \( \{E_j\} \) are the eigenvalues of \( H \), with the corresponding right eigenstates \( |\Psi^j_{R}\rangle \).
The obtained state satisfies $\rho_\tau=e^{-H\tau}\rho_0 e^{-H^\dagger\tau}$ and can be expanded as $\rho_\tau=\sum_{j,k} \frac{c_{jk}}{4}e^{-\tau(E_j+E_k^*)}\ket{\Psi_R^j}\!\bra{\Psi_R^k}$,
where $c_{jk}$ are coefficients (see Supplemental Material~\cite{SeeSM}).
Therefore, for a sufficiently long $\tau$,
the state with the smallest real-part eigenvalue dominates, projecting $\rho_0$ onto the right ground state $\ket{\Psi_R^s}$. Similarly, applying $U_\tau^\dagger=e^{-H^\dagger\tau}$ yields the left ground state $\ket{\Psi_L^s}$. In our experiment, we fix $\tau=10$ (in natural units).

Although direct implementation of non-unitary operations $U_{\tau}$ and $U_{\tau}^\dagger$ with gain is challenging for single photons, we bypass this obstacle by mapping them to passive evolutions without gain that share the same eigenstates~\cite{PhysRevLett.127.026404, PhysRevX.8.021017}.
Specifically, we factor out the global amplification to obtain $\tilde{U}_{\tau}=U_{\tau}e^{-\Lambda(\tau)}$, which shares the same eigenstates as $U_{\tau}$. Here, $\Lambda(\tau)=\ln\sqrt{\max_j|\lambda_j|}$ with $\{\lambda_j\}$ the eigenvalues of $U_{\tau}U_{\tau}^{\dagger}$.
Under this mapping, gain terms are converted to no loss, and loss terms to even greater loss, leading to a passive operation.

The passive operation ${\tilde{U}_{\tau}}$ is implemented via singular value decomposition \( \tilde{U}_{\tau} = VDW^{\rm T} \)~\cite{PhysRevX.8.021017}.
Here, $V$ and $W$ are unitary operations, and $D$ is a non-unitary diagonal matrix with the first diagonal element \( D_{11} = 1 \), while the remaining diagonal entries are less than 1.
We realize $D$ by introducing mode-selective photon losses using BDs and HWPs, where the loss strength is controlled by the HWPs between BD3 and BD4, while BD3 and BD4 separate and recombine the modes to implement mode selectivity.
The unitary operations $W$ and $V$ are further decomposed into a sequence of unitary operations, which are implemented by recombining the photon into a designated spatial mode with BDs and controlling the polarization with HWPs.
Finally, we reconstruct the output ground states $|\Psi^s_R\rangle$ and $|\Psi^s_L\rangle$ by quantum state tomography~\cite{SeeSM}.

Our setup provides a general and controllable method to implement imaginary-time evolution generated by an NH-PH in a photonic platform.
Enabling access to both the right and left ground states, it allows direct experimental studies of ground-state properties and their biorthogonal structure.
Moreover, the approach is not restricted to non-Hermitian Hamiltonians and can also be applied to Hermitian systems to prepare the corresponding ground states, making it broadly applicable for quantum simulation across condensed matter physics~\cite{Georgescu2014,Houck2012}, high-energy physics~\cite{Gerritsma2010}, and quantum chemistry~\cite{OMalley2016}.

\textit{Experimental results.}---
With the reconstructed left and right ground states of the NH-PH, we examine whether the experimentally obtained states faithfully reproduce the tailored properties by analyzing the expectation values of different order parameters $\langle O\rangle=\bra{\Psi_L^s}O\ket{\Psi_R^s}$.
To quantify non-reciprocal correlations, we choose \({O}_{\text{left}} ={\sigma}_1^{-} \otimes{\sigma}_{2}^{+}/2\) and \({O}_{\text{right}} = {\sigma}_1^{+} \otimes{\sigma}_{2}^{-}/2\), which probe spin-flip processes in opposite directions. Their difference, \({O}_{\text{chiral}} = {O}_{\text{right}} - {O}_{\text{left}}\), characterizes the chiral imbalance encoded in the underlying asymmetric valence bond---a designed asymmetry in the entanglement structure between the two sites.
Here, ${\sigma}^+ = {\sigma}^x + i {\sigma}^y$, ${\sigma}^- = {\sigma}^x - i {\sigma}^y$, and ${\sigma^{\alpha}}$ ($\alpha=x$, $y$, $z$) are the Pauli matrices.
For comparison, we also consider \({O}_{\text{AF}} = {\sigma}_1^z \otimes {\sigma}_{2}^z\),
which is commonly adopted to detect the conventional antiferromagnetic order~\cite{PhysRevLett.100.167202}.

The asymmetry parameter $\mu$ controls the non-Hermiticity of the system: $\mu=1$ corresponds to the Hermitian limit, while $\mu\neq1$ introduces non-Hermitian chirality. Theoretically, evaluating these order parameters on the designed ground states yields $\langle O_{\mathrm{AF}}\rangle=-1/4$, $\langle O_{\mathrm{left}}\rangle=-\mu/4$, and $\langle O_{\mathrm{right}}\rangle=-1/(4\mu)$. As shown in Fig.~\ref{fig2}(a), the experimentally obtained expectation values (denoted $\langle O^{\mathrm{H}}\rangle$, with error bars representing statistical uncertainties) agree well with these theoretical predictions across the entire range of $\mu$, faithfully reproducing the predicted $\mu$ dependence.
In the Hermitian limit ($\mu=1$), the two correlators $\langle O_{\text{left}}\rangle$ and $\langle O_{\text{right}}\rangle$ coincide (both equal to $-1/4$), and hence $\langle O_{\text{chiral}}\rangle=0$. When tuning $\mu$ away from $1$, $\langle O_{\text{chiral}}\rangle$ changes sign---positive for $\mu>1$ and negative for $\mu<1$---directly demonstrating the non-Hermitian chirality encoded in the MPS design.

Notably, despite the small system size ($N=2$), the observed trends and behaviors align closely with the theoretical predictions originally derived for infinitely large systems in Ref.~\cite{PhysRevLett.130.220401}. This agreement reflects the fact that the MPS-based NH-PH construction captures the essential bulk physics even at the minimal two-site level.

Beyond local observables, we further probe the internal structure of the reconstructed ground states $\ket{\Psi_{R}^{s}}$ and $\ket{\Psi_{L}^{s}}$ by examining their entanglement spectra $s_i$, defined via the Schmidt decomposition $\ket{\psi} = \sum_i s_i \ket{\psi_i^A} \otimes \ket{\psi_i^B}$~\cite{PhysRevB.81.064439}. As shown in Figs.~\ref{fig2}(b) and~\ref{fig2}(c), the entanglement spectra for both the right and left ground states exhibit a clear dependence on $\mu$: the two dominant Schmidt coefficients split asymmetrically as $\mu$ deviates from 1, with the degree of splitting increasing with $|\mu-1|$. This demonstrates that even in this minimal two-site setup, the NH-PH construction yields states with non-trivial internal entanglement that directly reflects the designed non-Hermitian asymmetry.

In summary, these results provide clear experimental evidence that the engineered NH-PH faithfully reproduces both the local and entanglement properties dictated by the designed biorthogonal MPS ground states. The agreement across multiple order parameters---probing non-reciprocal correlations, chiral imbalance, and antiferromagnetic order---along with the entanglement spectra, establishes a controlled photonic realization of NH-PHs with designed non-Hermitian properties.

\textit{Extension to a new $N=3$ model.}---
Having experimentally validated the NH-PH construction in the minimal $N=2$ realization, we extend the construction to a new spin-$1/2$ model with $N=3$ to explore an intrinsic non-Hermitian phase transition.
Here, the phase transition denotes that, once multiple local terms are present, the predesigned MPSs are not guaranteed to remain the global ground states due to the breakdown of the non-Hermitian variational principle~\cite{PhysRevLett.130.220401,Ashida02072020}, which is distinct from conventional phase transitions defined in the thermodynamic limit~\cite{Song2022,Beaulieu2025,Cai2021}.

\begin{figure}[!t]
    \centering
    \includegraphics[width=0.45\textwidth]{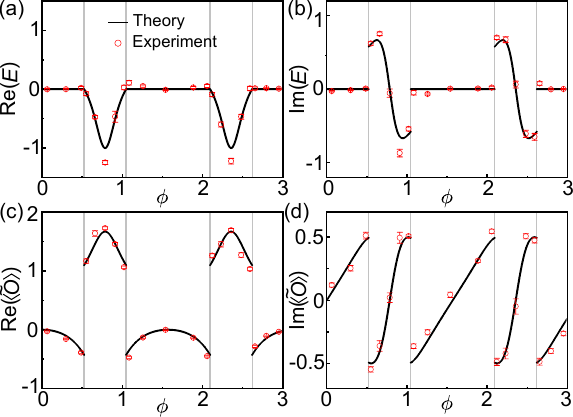}
    \caption{Real (a) and imaginary (b) parts of the ground-state energy of NH-PH $\tilde{H}$ ($N=3$).
    The real parts of eigenvalues are negative in $\pi/6<\phi<\pi/3$ and $2\pi/3<\phi<5\pi/6$.
    Real (c) and imaginary (d) parts of the expectation value of order parameter $\langle \tilde{O} \rangle=\bra{\Psi_{\tilde{L}}^{s}}\tilde{O}\ket{\Psi_{\tilde{R}}^{s}}/\braket{\Psi_{\tilde{L}}^{s}}{\Psi_{\tilde{R}}^{s}}$. Experimental data (dots) are obtained from states obtained via imaginary-time evolution, while theoretical results (solid black lines) are obtained from ground states calculated via exact diagonalization of the constructed NH-PH.
    Error bars indicate the statistical uncertainty, which are obtained by assuming Poissonian statistics.
    }
    \label{fig3}
\end{figure}

In the following, we consider an MPS with $d_v=2$, whose nonzero tensor elements contain
\begin{equation}
    [A^R]_{0 0}^{[0]} = e^{i \phi}, \quad
    [A^R]_{01}^{[1]} = [A^R]_{10}^{[1]} = 1.
\end{equation}
Here, $\phi$ is a tunable phase parameter in the local tensor.
These tensors define the right ground state $\ket{\tilde{R}}$.
The left ground state is taken as $\ket{\tilde{L}}=K\ket{\tilde{R}}$, where $K$ denotes the complex conjugation operator. Its tensor elements are given by $[A^{\tilde{L}}]^{[i]} = {[A^{\tilde{R}*}]^{[i]}}$, and $[A^{\tilde{R}*}]$ refers to the complex conjugate of $[A^{\tilde{R}}]$.
We construct the corresponding NH-PH $\tilde{H}$ from the chosen left/right MPSs under periodic boundary conditions~\cite{SeeSM} and implement it on a three-qubit photonic platform (two path qubits and a polarization qubit).

Extending the approach to implement the imaginary-time evolution for the $N=3$ model, we extract the corresponding right and left ground states $\ket{\Psi_{\tilde{R}}^{s}}$ and $\ket{\Psi_{\tilde{L}}^{s}}$ of $\tilde{H}$~\cite{SeeSM}.
We then evaluate the ground-state energy from $\bra{\Psi_{\tilde{L}}^{s}}\tilde{H}\ket{\Psi_{\tilde{R}}^{s}}$, whose real and imaginary parts are shown in Figs.~\ref{fig3}(a) and~\ref{fig3}(b), respectively.
The theoretical curves are obtained by directly calculating the right and left ground states of the constructed NH-PH and evaluating the corresponding ground-state energy.
As the parameter $\phi$ is varied, the real part of the ground-state energy becomes negative within two intervals $\phi \in (\pi/6, \pi/3)$ and $\phi \in (2\pi/3, 5\pi/6)$.
In these regions, the designated zero-energy MPS $\ket{\tilde{R}}$ $(\ket{\tilde{L}})$ is no longer the global right (left) ground state of the NH-PH, since new eigenstates $\ket{\Psi_{\tilde{R}}^{s}}$ ($\ket{\Psi_{\tilde{L}}^{s}}$) with lower negative energy emerge.
This signals an intrinsic non-Hermitian phase transition due to the breakdown of the variational principle in non-Hermitian systems~\cite{PhysRevLett.130.220401,Ashida02072020}.

Meanwhile, we can also capture this transition via the ferromagnetic order parameter $\bra{\Psi_{\tilde{L}}^{s}}\tilde{O}\ket{\Psi_{\tilde{R}}^{s}}$ with $\tilde{O}=\sigma^z_1 \otimes \sigma^z_2 \otimes \sigma^0_3$ [see Figs.~\ref{fig3}(c) and~\ref{fig3}(d)].
The expectation value exhibits abrupt jumps coinciding with the spectral crossings, confirming the presence of a phase transition~\cite{Alushi2024,Hotter2024}.
Thus, the $N=3$ case provides clear experimental evidence of a non-Hermitian phase transition arising from the fundamental difference between Hermitian and non-Hermitian parent Hamiltonians.
Moreover, we verify that the same phase transition remains for $N=4,6,10$~\cite{SeeSM}, supporting that this phase transition persists upon increasing the system size and is therefore not a finite-size effect.
It is worth noting that, different from conventional thermodynamic phase transitions, the present transition is inherently tied to the competition between complex eigenvalues. The conceptual connection between these level crossings and thermodynamic phase transitions in open systems remains an evolving area for future study.

\textit{Conclusion.}---
We have reported the first experimental realization of NH-PHs with tailored properties. Using single photons, we constructed NH-PHs for two models of different sizes---an $N=2$ system for proof-of-principle validation and an $N=3$ system demonstrating a non-Hermitian phase transition.

For the $N=2$ model based on asymmetric AKLT MPSs, we validated the NH-PH construction by measuring four order parameters probing non-reciprocal correlations, chiral imbalance, and antiferromagnetic order. The experimental results showed excellent agreement with theoretical predictions across the entire range of the asymmetry parameter $\mu$, confirming that the engineered Hamiltonian faithfully reproduces the designed biorthogonal ground states.

As a concrete application, we extended the scheme to an $N=3$ model and observed an intrinsic non-Hermitian phase transition. This transition is manifested by both a level crossing in the energy spectrum and an abrupt jump in the order parameter---a hallmark of non-Hermitian criticality distinct from Hermitian phase transitions. This behavior points to potential applications in enhanced sensing, where small perturbations near the critical point can be significantly amplified~\cite{Alushi2024,Hotter2024}.

In summary, our work not only enables systematic exploration of intrinsic non-Hermitian phases and phase transitions---such as the chiral phase demonstrated here---but also establishes a foundation for simulating non-Hermitian systems across diverse platforms, including photonics, cold atoms, and superconducting circuits. Looking forward, the NH-PH framework can be extended to study non-Hermitian topology, many-body dynamics, and quantum metrology in regimes beyond the reach of existing methods, opening new avenues for both fundamental physics and practical applications.


\begin{acknowledgments}{\it Acknowledgments.---}
This work has been supported by the National Key R\&D Program of China (Grant No.~2023YFA1406701) and National Natural Science Foundation of China (Grant Nos.~92265209, 12474352, 92476106, 12305008, 12374479).
YCG and SY are supported by the National Natural Science Foundation of China (NSFC) (Grant No.~12475022 and No.~125B2100) and the Quantum Science and Technology - National Science and Technology Major Project (Grant No.~2021ZD0302100).
KKW is supported by the Natural Science Foundation of Anhui Province (Grant No.~2508085Y002). LX and KKW also acknowledge support from the Open Research Fund of the Beijing National Laboratory for Condensed Matter Physics (Grant No.~2024BNLCMPKF010). DKQ and HXG acknowledge support from the National Postdoctoral Program for Innovative Talent (Grant Nos.~BX20230036, BX20250174) and the China Postdoctoral Science Foundation (Grant Nos.~2023M730198, 2024M760425).
\end{acknowledgments}

\EndMatterTitle

\begin{figure}[!t]
    \centering
    \includegraphics[width=0.45\textwidth]{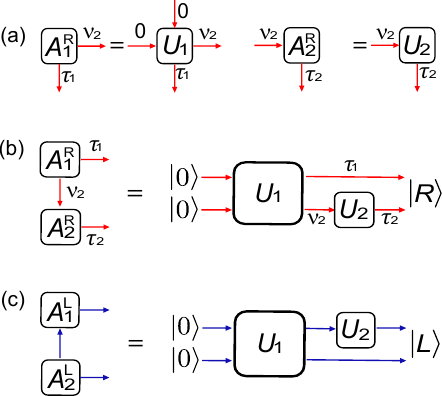}
    \vspace{1.5cm}
    \caption{(a) The mapping from MPSs tensors to unitary operations. Here, $U_1$ and $U_2$ are unitary operations defined in Eq.~(\ref{eq11}).
    (b-c) The quantum circuits mapped from the MPSs to generate $\ket{R}$ and $\ket{L}$. }
    \label{fig4}
\end{figure}

\noindent
\textit{Direct MPS preparation for $N=2$.}---
As an independent cross-check, we directly generate the target AKLT states $\ket{R}$ and $\ket{L}$ by mapping the MPS tensors onto a quantum circuit.
If the NH-PH approach works, the steady states extracted from imaginary-time evolution satisfy $\ket{\Psi_R^s}=\ket{R}$ and $\ket{\Psi_L^s}=\ket{L}$.
This mapping is constructed from the right-canonical form of the MPS~\cite{PerezGarcia2007}.
\begin{equation}
[A_1^R]_{\nu_2}^{\tau_1} = \frac{1}{\sqrt{\mu^2 + 1}}
\begin{pmatrix}
1 & 0 \\
0 & \mu
\end{pmatrix}, \quad
[A_2^R]_{\nu_2}^{\tau_2} =
\begin{pmatrix}
0 & -1 \\
1 & 0
\end{pmatrix}.
\label{eq7}
\end{equation}
Here, the rows represent the physical indices and the columns correspond to the virtual indices spanned by~\mbox{$\bigl\{|0), |1)\bigr\}$}.
Specifically, each local tensor with bond dimension \(d_v\) and physical dimension \(d_p\) is converted to a \(\mathrm{SU}(d_p d_v)\) unitary matrix~\cite{Barratt2021, SeeSM}.
Taking the boundary condition into account, each element of two local tensors $[A_j^R]$ can be directly transformed into the corresponding $U_j$ [see Fig.~\ref{fig4}(a)]
\begin{equation}
\begin{aligned}
    &[U_1]_{(\tau_1\otimes\nu_2), ([0]\otimes[0])} = [A_1^R]^{\tau_1}_{\nu_2},\\
    &[U_2]_{\tau_2, \nu_2} = [A_2^R]^{\tau_2}_{\nu_2},\\
\end{aligned}
\end{equation}
from which we obtain the unitary operations in the circuit as [see Fig.~\ref{fig4}(b)]
\begin{equation}
    U_1 =
\begin{pmatrix}
\frac{1}{\sqrt{\mu^2 + 1}} & 0 & 0 & -\frac{\mu}{\sqrt{\mu^2 + 1}} \\
0 & 1 & 0 & 0 \\
0 & 0 & 1 & 0 \\
\frac{\mu}{\sqrt{\mu^2 + 1}} & 0 & 0 & \frac{1}{\sqrt{\mu^2 + 1}}
\end{pmatrix},
\quad
U_2 =
\begin{pmatrix}
0 & -1 \\
1 & 0
\end{pmatrix}.
\label{eq11}
\end{equation}
The left ground state \(|L\rangle = \mathcal{P} |R\rangle\) can be obtained by exchanging two qubits [see Fig.~\ref{fig4}(c)].

\begin{figure}[!t]
    \centering
    \vspace{1.5cm}
    \includegraphics[width=0.47\textwidth]{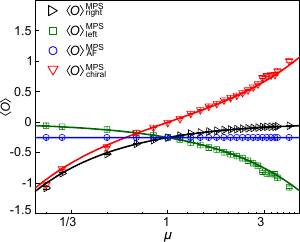}
    \caption{The expectation values of order parameters $\langle O \rangle=\langle L|{O}|R\rangle / \langle L|R\rangle$ for NH-PH $H$ with $N=2$.
        $\langle O^{\text{MPS}} \rangle$ denotes the results from direct measurements in the quantum circuits.}
    \label{fig5}
\end{figure}

We experimentally implement these two quantum circuits to prepare $\ket{R}$ and $\ket{L}$ using single photons.
The photons are first initialized in horizontal polarization by passing through a PBS.
A BD is inserted to introduce two spatial modes, namely the upper and lower paths.
In each path, the photons are prepared in state $\ket{r}\ket{h}$, and then we implement the circuits $(\mathbb{I}\otimes U_2) U_1$ and $(U_2\otimes \mathbb{I}) U_1$ on the upper and lower paths, respectively.
The unitary operations $U_1$ and $U_2$ are implemented with BDs and HWPs, where the relevant parameters are controlled by the setting angles of the HWPs~\cite{SeeSM}.
The expectation values $\bra{L} O \ket{R}$ are obtained by two steps:
(i) treating the order parameter $O$ as operations in the upper paths to produce $O | R \rangle$.
(ii) vertically placing BD to recombine the upper and lower paths and measure $\langle O \rangle$ using interferometric measurements.

Figure~\ref{fig5} summarizes the order parameters measured from the directly prepared MPS states.
The measured $\langle O^{\mathrm{MPS}}\rangle$ agrees with the theoretical prediction (solid lines) and matches the values extracted from the imaginary-time evolution in Fig.~\ref{fig2}(a).
This direct measurement of the expectation values of order parameters provides a cross-check of our experiment and confirms that the steady state produced by imaginary-time evolution faithfully represents the target MPSs.

From the technical perspective, mapping MPSs to quantum circuits is a general approach~\cite{Barratt2021}.
Our experiment provides a practical route for the experimental preparation of MPS states and the direct measurement of expectation values of observables.
The same strategy can be extended to larger $N$ systems, providing a new approach to quantum simulation of many-body systems.

\end{document}